\documentstyle[12pt]{article}
\input epsf

\newlength{\dinwidth}
\newlength{\dinmargin}
\newcommand{\resection}[1]{\setcounter{equation}{0}\section{#1}}

\begin{document}

\begin{center}
  \begin{Large}
  \begin{bf}
STUDY OF THE ANOMALOUS COUPLINGS AT NLC WITH POLARIZED BEAMS
\\
  \end{bf}
  \end{Large}

    \vspace{5mm}

  \begin{large}
R. Casalbuoni$^{(a,b)}$,  S. De Curtis$^{(b)}$ and D. Guetta$^{(a,c)}$ \\
  \end{large}
  \vspace{5mm}
\begin{small}
$(a)$ - {\it Dipartimento di Fisica, Universit\`a di Firenze, Italy}\\ $(b)$ -
{\it I.N.F.N., Sezione di Firenze, Italy}\\ $(c)$ - {\it Department of Physics,
Technion, Haifa, Israel}.\\
\end{small}
  \vspace{5mm}
LC note: LC-TH-1999-013

\end{center}

\setcounter{page}{1}
\newcommand{\ac}{\`{a} \  }
\newcommand{\ec}{\`{e} \  }
\newcommand{\be}{\begin{equation}}
\newcommand{\ee}{\end{equation}}
\newcommand{\bea}{\begin{eqnarray}}
\newcommand{\eea}{\end{eqnarray}}
\newcommand{\W}{\tilde{W} }
\newcommand{\Y}{\tilde{Y} }
\newcommand{\et}{\tilde{e} }
\newcommand{\M}{\tilde{M} }
\newcommand{\g}{\tilde{g} }
\newcommand{\pli}{\tilde{\psi}_{L} }
\newcommand{\pri}{\tilde{\psi}_{R} }
\newcommand{\bW}{{\bf \tilde{W}} }
 \newcommand{\bV}{{\bf V} }
 \newcommand{\bY}{{\bf \tilde{Y}} }
 \newcommand{\st}{\tilde{s}_{\theta} }
 \newcommand{\ct}{\tilde{c}_{\theta} }
 \newcommand{\s}{s_{\theta} }
  \newcommand{\kt}{c_{\theta} }
  \newcommand{\kdt}{c_{2\theta} }
 \newcommand{\A}{\tilde{A} }
 \newcommand{\Z}{\tilde{Z} }
 \newcommand{\q}{(\frac{g}{g^{\prime\prime}})^{2}}

  \newcommand{\epem}{e^{+}e^{-}}
  \newcommand{\sd}{s_{2\theta} }
  \newcommand{\te}{t_{\theta} }
 \newcommand{\lnab}{ieg_{_{ZWW}}[g^{\nu\alpha}(-2p_{3}-p_{4})^{\beta}
 +g^{\alpha\beta} (p_{3}-p_{4})^{\nu}+g^{\beta\nu}(2p_{4}+p_{3})^{\alpha}]}
 \newcommand{\abzq}{(a_{e}^{Z})^{2}+(b_{e}^{Z})^{2}}
 \newcommand{\abz}{a_{e}^{Z}+b_{e}^{Z}}
 \newcommand{\aez}{a_{e}^{Z}}
 \newcommand{\bez}{b_{e}^{Z}}
\newcommand{\Pe}{P_L(e^-)}
\newcommand{\Pep}{P_L(e^+)}

\newcommand{\nn}{\nonumber}
\newcommand{\dd}{\displaystyle}
\newcommand{\AAA}{\frac{e^{2}}{4} \sin\phi}
\newcommand{\AAB}{\frac{e^{2}}{4\sqrt{2}} (\cos\phi-1)}
\newcommand{\AAC}{\frac{e^{2}}{4\sqrt{2}} (\cos\phi+1)}
\resection{Introduction}

The experimental confirmation of the Standard Model (SM) is presently limited to
the sector of fermion interactions with gauge vector bosons, however a precise
measure of the vector boson self coupling is needed to test the non abelian
structure of the SM. In fact indirect tests via loop effects must be confirmed
by direct exploration of the interactions of the vector bosons among themselves
via production and scattering experiments. Next linear colliders will provide
the right framework where to study the  $SU(2)_L\otimes U(1)_Y$ gauge structure
\cite{murayama}. Future experiments at these colliders may be used to confirm
the SM predictions as well as to probe for new physics. In particular it is
possible that signals  of new physics will appear in the gauge boson sector
through the discovery of anomalous vertices.
\newline\indent
Among the various possible reactions, where to
test the trilinear gauge boson couplings  we
choose the reaction $e
^+e^-\to W^+W^-$ \cite{hagiwara} with the $W$-pair
decaying into a lepton pair plus jets. We consider possible deviations from the
SM coming both from anomalous gauge vector boson trilinear couplings and from
non standard couplings of gauge bosons to fermions \cite{Dafne}.  Both of them
 can be due to the physics responsible of the electroweak
symmetry breaking.  Moreover we will see that the
trilinear and the fermionic couplings are strongly
correlated, therefore it is not justified to
neglect the latter in a phenomenological analysis.
\newline\indent
We assume the experimental possibility of reconstructing the polarization of the
final $W$'s and we consider the effects of having both the electron and the
positron beams polarized \cite{polarization}. This would be of particular
interest in order to increase the sensitivity to the anomalous couplings
\cite{paver}. The observables on which we concentrate to bound new physics are
the differential cross sections for the $e^+e^-\to W^+W^-$ for the various
channels of final $W$ and initial beams polarization. Theoretically, this
program is carried out by evaluating the helicity amplitudes
$F^{\lambda\lambda^\prime}_{\tau\tau^\prime}$, where $\lambda$
($\lambda^\prime$) are the $e^-$ ($e^+$) helicities
($\lambda^\prime=-\lambda=\pm 1/2$), and $\tau$ ($\tau^\prime$) = $\pm 1, 0$ are
the $W^-$ ($W^+$) helicities. From the study of the high-energy behaviour of the
observables \cite{Dafne} we concluded that terms proportional to the square of
the anomalous couplings are not negligible and then it is not justified to make
any approximation in the deviations from the SM. For this  reason we have used
the exact expressions in our calculation. We analyze various models at energy
ranges relevant to next linear colliders and for various degree of longitudinal
polarization of the incoming beams.
\newline\indent
In Section 2 we will  write the general parametrization of the gauge boson
trilinear anomalous couplings and of the non-standard fermionic couplings and we
will introduce the observables considered in our analysis.   The
phenomenological analysis is given in  Section 3. Although a general fit could
be possible, we have limited our analysis to the case of two free parameters by
 considering  particular models. The constraints on the anomalous couplings from
future $e^+e^-$ colliders  are shown in Section 4. The options we will consider
are for
 c.m. energies of 300, 500 and 1000 $GeV$, with corresponding integrated
luminosities of 200, 300 and 500 ${\rm fb}^{-1}$. In this section we will also
study the effect of the polarization of the electron and positron beams and we
will see that  a more restricted region for the anomalous parameters can be
obtained. Conclusions and comparison with the reaches of other collider machines
are given in Sect. 5.

\resection{Helicity amplitudes and polarized cross sections}

The $CP$ invariant effective Lagrangian for the
gauge boson self-interactions in terms of the
anomalous couplings can be expressed as follows:
\cite{effective}
\bea
\label{leff}
{\cal L}^{3}_{eff} & = &
   -ie[A_{\mu}(W^{-\mu\nu}W^{+}_{\nu} - W^{+\mu\nu}W^{-}_{\nu})+
   A^{\mu\nu}W^{+}_{\mu}W^{-}_{\nu} ]-
   iex_{\gamma }A^{\mu\nu}W^{+}_{\mu}W^{-}_{\nu} \nonumber \\
   &   & -ie(ctg\theta +\delta _{Z} )
   [Z_{\mu}(W^{-\mu\nu}W^{+}_{\nu}-W^{+\mu\nu}W^{-}_{\nu})]-
   iex_{Z} Z^{\mu\nu}W_{\mu}^{+}W_{\nu}^{-} \nonumber \\
   &   & +ie\frac{y_{\gamma}}{M_{W}^{2}}A^{\nu\lambda}W^{-}_{\lambda\mu}
   W^{+\mu}_{\nu} +ie\frac{y_{Z}}{M_{W}^{2}}Z^{\nu\lambda }W^{-}_{\lambda\mu}
   W^{+\mu}_{\nu}  \label{lan}
\eea
where $V_{\mu\nu}=\partial_\mu V_\nu-\partial_\nu V_\mu$ for $V=A,~Z,~W$.
$x_\gamma$ and $x_Z$ parametrize deviations in weak and electromagnetic dipole
coupling from their SM values; $y_\gamma$ and $y_Z$ indicate the intensity of
the non-standard quadrupole interactions of $W^{\pm}$ and $\delta_Z$ describes a
deviation  overall of the  $ZW^+W^-$ coupling from its standard value. The
relations of the above  parameters to those directly connected with $W$ static
properties are (se e.g. \cite{gouna})
\be
\Delta g_1^Z=\delta_Z\tan\theta,~~~~
\Delta \kappa_\gamma=x_\gamma,~~~~
\lambda_\gamma=y_\gamma
\ee
\be
\Delta \kappa_Z=(x_Z+\delta_Z)\tan\theta,~~~~~\lambda_Z=y_Z\tan\theta
\label{g1z}
\ee
We will assume that the energy relevant to the problem (i.e. the energy of the
electron-positron beams) is much lower than the energy scale of the new physics.
As a consequence we will consider all the anomalous couplings as constant
ignoring their possible energy dependence.
\newline\indent
The fermionic Lagrangian relevant to our process is:
 \be
 {\cal L}^{neutral} =
  -eZ_{\mu}\bar{\psi}_e
  [\gamma^{\mu}a_{e}^{Z}-\gamma^{\mu}\gamma_{5}b_{e}^{Z}]\psi_e
  -eA_{\mu}\bar{\psi}_e
  Q_{e}\psi_e\label{Lbn}
  \ee
  \be
  {\cal L}^{charged}= e  a_{w}\left(
 \sum_{\ell}\bar{\psi}_{\nu_{\ell}}\gamma^{\mu}(1-\gamma_{5})
  \psi_{\ell}W^{+}_{\mu} +
  \sum_{i=1}^3\bar{\psi}_{i}\gamma^{\mu}(1-\gamma_{5})\tau_+
\psi_{i}W^{+}_{\mu}+ h.c. \right)
\ee
In these expressions $\psi_e$ is the
electron field, $\ell=e,\mu,\tau$, and the sum over $i=1,2,3$ is on the quark
charged current eigenstates. In   the following we will consider the possibility
of deviations of  the parameters $a_e^Z$, $b_e^Z$ and $a_W$ from their SM
values. We define
\be
  f=f(SM)+\delta f \ee
   with $f=a_{e}^{Z},b_{e}^{Z},a_W$,  and
\be  a_{e}^{Z}(SM)=
\frac 1 {2\s\kt}(-\frac{1}{2}+2\s ^{2}),~~~~
b_{e}^{Z}(SM) =  -\frac{1}{4\s\kt},~~~~
a_{W}(SM) = \frac{1}{2\sqrt{2}\s}
\ee
where $\s$ is defined through the input parameters $G_F$, $\alpha(M_Z)$, $M_Z$:
\be
\s^2=\frac 1 2-\sqrt{\frac 1 4-\frac{\pi\alpha(M_Z)}{\sqrt{2}G_F M_Z^2}}
\ee
Finally we put
\be
M_W=M_W(SM)+\delta M_W,~~~~~~~M_W(SM)=\kt M_Z
\ee
The differential cross-section for initial $e^+_{\lambda^\prime}$,
$e^-_\lambda$, and final $W^+_{\tau^\prime}$, $W^-_\tau$, where $\lambda$,
$\lambda^\prime$, $\tau$ and $\tau^\prime$ are the helicities of the various
particles, is given by
\be
       \frac{d\sigma ^{\lambda \lambda ^{\prime}}_{\tau \tau^{\prime}}}
           {d\cos\phi}=\frac{|\vec{p}|}{4\pi s \sqrt{s}}
           |F^{\lambda\lambda^{\prime}}_{\tau\tau^{\prime}}(s,\cos\phi)|
           ^{2}
\ee
where
\be
|\vec p|=\beta_W\frac{\sqrt{s}} 2,~~~~~\beta_W=\sqrt{1-\frac{4M_W^2}{ s}}
\ee
and $\phi$ is the angle between the incoming electron and $W^-$.
\newline
 The expression of the helicity amplitudes in terms of the anomalous couplings
are given in the first of ref. \cite{Dafne}. In this paper we give the helicity
amplitudes in a form where the contribution of the SM and of the anomalous part
is explicit
\be
F^{\lambda\lambda^{\prime}}_{\tau\tau^{\prime}}=
F^{\lambda\lambda^{\prime}}_{\tau\tau^{\prime}}(SM)+
\delta F^{\lambda\lambda^{\prime}}_{\tau\tau^{\prime}}
\ee
In the helicity basis the transition probability is given by \cite{paver2}:
\be
|F|^2=\frac{1}{4}\left[\left(1-P_L(e^-)\right)\left(1+P_L(e^+)\right)|F_{-}|^2+
\left(1+P_L(e^-)\right)\left(1-P_L(e^+)\right)|F_{+}|^2\right]
\ee
where $F_{\pm}$ correspond to $\lambda=-\lambda^{\prime}=\pm 1/2 $ and
$P_L(e^-)(P_L(e^+))$ is the degree of longitudinal polarization of the
$e^{-}(e^{+})$ beam. Then the differential cross-sections are given as
\cite{paver2}
\bea
\frac{d\sigma_{\tau\tau^\prime}}{d\cos\phi}&=&
\frac 1 4 \left[(1+P_L(e^-))(1-P_L(e^+))
\frac{d\sigma^{(+1/2,-1/2)}_{\tau\tau^\prime}}{d\cos\phi}
\right.
\nonumber \\
 & &~~~~+
\left.
(1-P_L(e^-))(1+P_L(e^+))\frac{d\sigma^{(-1/2,+1/2)}_{\tau\tau^\prime}}
{d\cos\phi}\right]
\eea
We consider the experimental possibilities of reconstructing the polarization of
the final $W^+ W^-$ and then we have three different differential cross sections
as  observables:
\be
\frac{d\sigma_{TT}}{d\cos\phi};~~
\frac{d\sigma_{LL}}{d\cos\phi};~~
\frac{d\sigma_{LT}}{d\cos\phi}~ .
\ee
For the standard $SU(2)_L\otimes U(1)_Y$ couplings the asymptotic behaviour in
the high-energy limit of these observables is:
\be
\frac{d\sigma_{TT}}{d\cos\phi}\sim \frac{1}{s};~~
\frac{d\sigma_{LL}}{d\cos\phi}\sim\frac{1}{s};~~
\frac{d\sigma_{LT}}{d\cos\phi}\sim\frac{1}{s^2}~.
\ee
In fact within the SM the $e^+e^-\to W^+W^-$ cross-section has a good behaviour
at high-energy due to the cancellations required by the gauge invariance. In the
actual case, however, this is not true. However we should remember that we are
dealing with an effective theory and that we are allowed to use it only up to
energies smaller than the new physics scale. The complete analysis of the
high-energy behaviour of the helicity amplitudes in presence of the anomalous
couplings is given in \cite{Dafne}.  The main result is that the terms linear in
$s$, which  are the most important ones at high-energy, turn out to be quadratic
in the anomalous couplings. Since the expansion parameter is $s/M_W^2$, it is a
bad approximation to neglect the quadratic terms in the anomalous parameters
when one is discussing the cross-sections at energies of the order $10~M_W$ or
more (we are assuming a size for the anomalous parameters comparable to the size
of the radiative corrections within the SM). From the explicit expressions in
 \cite{Dafne} it turns out that the contribution of the dipole interactions
$x_\gamma,x_Z\neq 0$ and the deviation from universality $\delta_Z\neq 0$ lead
to $\sigma_{LL}\sim s$; quadrupole interactions $y_\gamma,y_Z\neq 0$ imply that
$\sigma_{TT}\sim s$  and that the leading term in the high-energy expansion of
$d\sigma_{LT}$ depend on $x_\gamma+y_\gamma$, $2\delta_Z+x_Z+y_Z$, $\delta
a_e^Z$, $\delta b_e^Z$, $\delta a_W$.  This study suggests that an analysis  of
the differential cross-sections at various  energies should be useful in order
to discriminate among various anomalous parameters.

\resection{Phenomenological analysis}

As a procedure to quantitatively assess the sensitivity of the differential
cross sections  to the anomalous gauge couplings and to the non-standard
fermionic couplings, we divide the experimentally significant range of the
production angle $\cos\phi$ ($-0.95\le\cos\phi\le 0.95$) into 6 bins. For each
differential cross-section we  evaluate the differences
\be
\sigma_{i,a}^{AN}=\sigma_{i,a}-\sigma_{i,a}(SM),~~~~i=1,\cdots,6,~~~a=
LL,LT,TT
\ee
where $\sigma_{i,a}$ and $\sigma_{i,a}(SM)$ are respectively the full
and the SM differential cross-sections integrated over the bin $i$. From
these differences we can evaluate the $\chi^2$ function for each differential
cross-section
\be
\chi^2_a=\sum_{i=1}^6\frac{(\sigma_{i,a}^{AN})^2}{(\delta\sigma_{i,a})^2},
~~~~~~a=LL,LT,TT
\label{chi2}
\ee
We assume the errors $\delta\sigma_{i,a}$ as follows
\be
\delta\sigma_{i,a}=\sqrt{\frac{\sigma_{i,a}}{BR\cdot L}+
\sigma_{i,a}^2 \Delta_{\rm sys}^2}
\label{err}
\ee
where $\Delta_{\rm sys}$ is the systematic error taking into account the
relative errors on the luminosity, on the branching ratio of the decay of $W$'s
into a lepton pair plus jets, on the acceptance and on the longitudinal
polarization of the  beams. We assume
 $\Delta_{\rm sys}=1.5\%$. We take into account the efficiency in reconstructing
the $W$ pairs from their decays into a lepton pair plus jets by reducing the
true branching ratio from 0.29 to 0.10 (see \cite{BR}). This is the effective
branching ratio $BR$ that we use in eq. (\ref{err}). There is a delicate point
about the branching ratio. In fact we are assuming that the coupling $We\nu$ may
differ from its SM value. This could modify the branching ratios of $W$'s into
fermions. However, for deviations respecting the universality, it is easy to see
that neglecting, as usual, mixed terms in the radiative and in the anomalous
corrections, the branching ratios are the same as in the SM up to 1-loop level.

Let us now consider the bounds on the anomalous couplings coming from the
minimization of the $\chi^2$-function given in eq. (\ref{chi2}).  We will not
attempt to make a general analysis in all the parameters expressing the
deviations from the SM (5 anomalous trilinear couplings,  3 non-standard
fermionic couplings and the $M_W$ deviation), but rather we will concentrate on
a few models depending only on 2 parameters. The first simplifying hypothesis we
will make is that all the fermionic deviations can be expressed in terms of the
$\epsilon_i$, $i=1,2,3$, parameters \cite{altarelli}. By using the definitions
of the $\epsilon_i$ in terms of the observables one gets
\bea
\delta a_e^Z&=&-\frac {\epsilon_1}{8\s\kt}-\frac 1
{\kdt}\frac{\s}{\kt}\left(\frac{\epsilon_1} 2-\epsilon_3\right)\nn\\
\delta b_e^Z&=&\frac {\epsilon_1}{8\s\kt}\nn\\
\delta a_W&=&\frac 1{2\sqrt{2}\s}\left(\frac{\kt^2}{2
\kdt}\epsilon_1-\frac{\epsilon_2} 2-\frac{\s^2}{\kdt}\epsilon_3\right)
\nn\\
\frac{\delta
M_W^2}{M_W^2}&=&\frac{\kt^2}{\kdt}\epsilon_1-\epsilon_2-2\frac{\s^2}{\kdt}\epsilon_3
\eea
We will restrict our analysis to models  of new physics with no isospin
violation. In such a case one has $\epsilon_1=\epsilon_2=0$.  Therefore all the
non-standard fermionic couplings and $\delta M_W$ are parametrized in terms of
$\epsilon_3$. The experimental bound on $\epsilon_3$ coming from LEP1, LEP2,
Tevatron and low-energy  measurements is \cite{HEPconf}
\be
\epsilon_3^{\rm exp}=(3.55\pm 0.96)\times 10^{-3}
\ee
Notice that the $\epsilon_3$ appearing in our equations is only the contribution
due to new physics, so in order to get an useful bound for it one has to
subtract from $\epsilon_3^{\rm exp}$ the contribution from the SM radiative
corrections which, for  $m_{top}=175~GeV$, are the following \cite{cara}
\bea
m_H=100~GeV~~~~~&&\epsilon_3^{\rm rad}=5.11\times 10^{-3}\nn\\
m_H=300~GeV~~~~~&&\epsilon_3^{\rm rad}=6.115\times 10^{-3}\nn\\
m_H=1000~GeV~~~~~&&\epsilon_3^{\rm rad}=6.65\times 10^{-3}
\eea
The 90\% C.L. bound we get (in the case of two degrees of freedom) is then
\be
-(4.6_{+0.5}^{-1.0})\times
10^{-3}\le\epsilon_3\le(-0.5_{-0.5}^{+1.0})\times 10^{-3}
\label{eps}
\ee
The upper and lower bounds correspond to $m_{H}=100~GeV$ and $m_{H}=1~TeV$
respectively. The central value is obtained for $m_{H}= 300~GeV$.
\newline\indent
As far as the anomalous trilinear gauge couplings are concerned we will also
consider some particular  models. In order to reduce the number of free
parameters we will examine the Lagrangian (\ref{lan}) in terms of its underlying
symmetries.

\noindent\underline{MODEL A}\\
\noindent
First we consider the case in which one assumes a global $SU(2)_L$ symmetry for
the Lagrangian (\ref{leff}) \cite{bilenky}. This gives rise to the following
correlations between the anomalous parameters
\be
\delta_{Z}=x_{\gamma}=x_{Z}=0
\ee
and
\be
y_Z=\frac{\kt}{\s}y_\gamma\neq 0,
\ee
Assuming again $\epsilon_3\not=0$ we get a two parameter space
$(\epsilon_3,y_\gamma)$.

\noindent
\underline{MODEL B}\\
\noindent
 This is an effective model in which the existence of a new triplet of vector
fields ${\vec V}_\mu$ is assumed (BESS model  \cite{BESS}). These are the gauge
fields associated to a spontaneously broken local symmetry $SU(2)_V$. The new
vector particles mix with ${W}$ and $Z$. As a consequence at energies well below
their mass (assumed to be around the $TeV$ scale), effective anomalous fermionic
and trilinear couplings are generated \cite{lowen}. The parameters
characterizing this model are the gauge coupling of the ${\vec V}_\mu$,
$g^{\prime\prime}$, and their direct coupling to the fermions, $b$. Only
$\epsilon_3$ and $\delta_Z$ are different from zero and they are given by
\be
\epsilon_3=-\frac b 2+(\frac g {g^{\prime\prime}})^2,
~~~~
\delta_Z=\frac{\kt}{2\s\kdt}(b-\frac 1{\kt^2}
(\frac g {g^{\prime\prime}})^2)
\ee
where $g$ is the standard coupling of $SU(2)_L$. The BESS model is a
generalization of a non-linear $\sigma$-model and it is not renormalizable.
However at 1-loop level it is equivalent to a heavy Higgs model with the Higgs
mass playing the role of a cut-off. Assuming the cut-off of the order of 1 $TeV$
we can evaluate $\epsilon_3^{\rm rad}$ with a Higgs mass equal to 1 $TeV$. In
Fig. 1 we give the bounds obtained in this way on the parameter space arising
from LEP1, LEP2, Tevatron and low-energy experiments.
\begin{figure}
\epsfxsize=8truecm
\centerline{\epsffile[50 263 506 693]{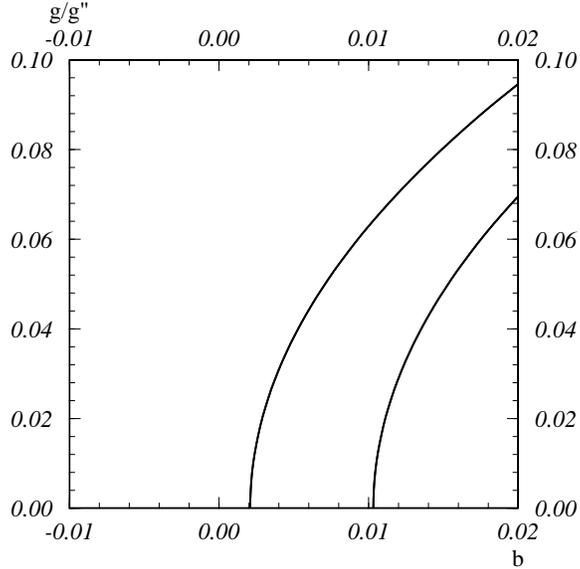}}
\caption{90\% CL bounds on the parameter space $(b,
g/g^{\prime\prime})$ in the case of model B (BESS model), as
obtained from the experimental data fit to the $\epsilon_3$ parameter.}
\end{figure}

\noindent
\underline{MODEL C}\\
\noindent
Finally we consider a model with two trilinear anomalous couplings and no
fermionic one. In particular we consider the case in which one adds to the SM
Lagrangian the most general operator of dimension 6 invariant under
$SU(2)_L\otimes U(1)_Y$ (see ref. \cite{bilenky}). The general form of this term
is
\be
\frac{ie}{2M_W^2}\left[\frac{f_B}{\kt} B_{\mu\nu}
\left(D_\mu\phi\right)^\dagger\left(D_\nu\phi\right)+
\frac{f_w}{\s} {\vec w}_{\mu\nu}\cdot
\left(D_\mu\phi\right)^\dagger\vec\tau\left(D_\nu\phi\right)\right]
\label{fbfw}
\ee
where $B_{\mu\nu}$ is the weak hypercharge field strength, $\vec w_{\mu\nu}$ is
the non-abelian field strength associated to ${\vec W}_\mu$ and $D_\mu$ the
covariant derivative operating on the Higgs field $\phi$. The interaction
generated by this operator is non-renormalizable but the symmetry of the
non-standard interaction (\ref{fbfw}) at one-loop level is expected to assure
logarithmic divergences only. For example for $f_B=fw$ one obtains the HISZ
model \cite{HISZ} analyzed at the NLC in \cite{barklow}. Here we  consider the
case $f_B=0$ leading to the following relations
\be
x_Z=-\frac{\s}{\kt}x_\gamma,~~~\delta_Z=\frac{1}{\s\kt}x_\gamma,~~~
y_Z=\frac{\kt}{\s}y_\gamma
\ee
Assuming  $\epsilon_3=0$ we get a two parameter space
$(x_\gamma,y_\gamma)$.

\resection{Bounds on the anomalous couplings}

We first analyze the models for various observables, namely the  unpolarized
cross-sections for different final $W$'s polarization, $\sigma_{LL}$,
$\sigma_{LT}$, $\sigma_{TT}$ and the total one.
\newline\indent
The 90\% CL bounds for the model A are given in Fig. 2, together with the bounds
coming from combining the $\chi^2$ for the final state polarized observables
($\sigma_{\rm comb}$), for a NLC machine with a center of mass energy of 500
$GeV$ and an integrated luminosity of 300 $fb^{-1}$. The figure shows the
independence of $\sigma_{LL}$ on $y_\gamma.$ We notice that we can obtain a good
restriction in the parameter space just combining the total $\sigma$ with
$\sigma_{LL}.$ In Fig. 3 we give the 90\% CL bounds arising from  $\sigma_{\rm
comb}$ at various energies and luminosities, namely: $E=300,~500,~1000~GeV$ with
$L=200,~300,~500~fb^{-1}$ respectively. We see that, for this model,  one needs
to reach at least 500 $GeV$ in energy in order to get bounds on $\epsilon_3$
better that the present ones (see eq. (\ref{eps})). Notice that the bound at
$E=1000~GeV$ includes  a region which is not connected with the SM point. This
means that there is the possibility of having an ambiguity at this energy.
However this is already resolved by the present data. It is also interesting to
notice that there are strong correlations among $y_\gamma$ and $\epsilon_3$.
This means that going at values of $\epsilon_3$ different from zero, the bounds
on $y_\gamma$ can change considerably.
\begin{figure}
\epsfxsize=8truecm
\centerline{\epsffile[50 263 506 693]{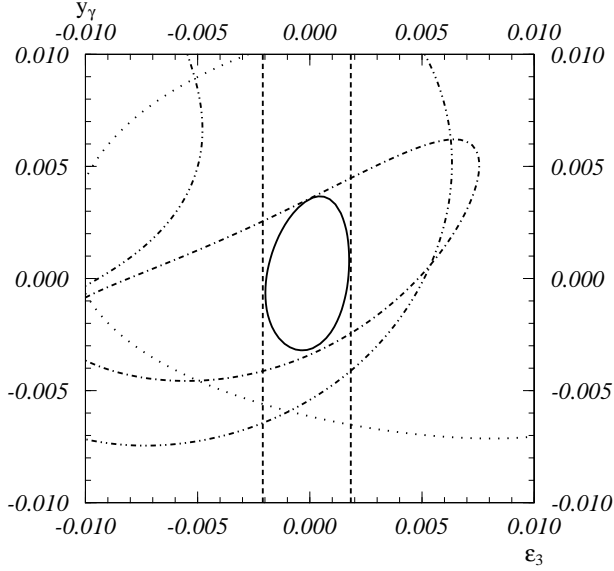}}
\caption{
90\% CL bounds on the parameter space $(\epsilon_3,y_\gamma)$ in the case of
model A. The dot line represents the bounds from $\sigma_{TT}$, the dash-dot
from $\sigma_{LT}$, the dash from $\sigma_{LL}$, the dash-doubledot  from
$\sigma_{\rm total}$, and the continuous one  from $\sigma_{\rm comb}$. Here the
parameters of the NLC are $E=500~GeV$,  $L=300~fb^{-1}$, and $\Pe=\Pep=0$.}
\end{figure}
\begin{figure}
\epsfxsize=8truecm
\centerline{\epsffile[50 263 506 693]{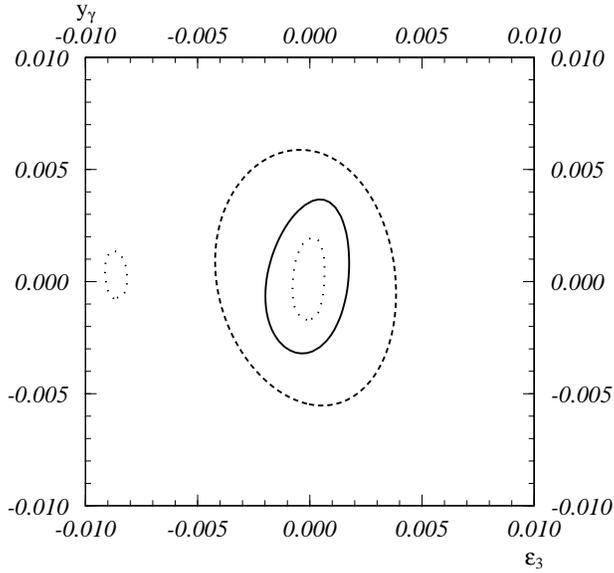}}
\caption{90\% CL bounds on the parameter space $(\epsilon_3,y_\gamma)$
in the case of model A. The bounds are obtained from $\sigma_{\rm comb}$  and
$\Pe=\Pep=0$ at various energies and luminosities. The dash line corresponds to
$E=300~GeV$ and $L=200~fb^{-1}$, the solid to $E=500~GeV$ and $L=300~fb^{-1}$,
and the dot one  to $E=1000~GeV$, $L=500~fb^{-1}$.}
\end{figure}
\newline\indent
The analysis at the NLC of the bounds on the BESS model parameter space at fixed
energy and different observables shows that the most restrictive ones comes from
$\sigma_{LL}$. In fact the new resonances are strongly coupled to $W_LW_L$
leading to an enhancement in this channel. In Fig. 4 we give the 90\% CL bounds
from $\sigma_{\rm comb}$ at various energies for unpolarized beams. The
comparison with Fig. 1 shows that already at 500 $GeV$ with the high luminosity
option we are considering, we can achieve a real improvement. Fig. 5 is the
analogous of Fig. 3 for the model C. We see that,  even for unpolarized beams,
the NLC configuration corresponding to an energy of 500 $GeV$ and
$L=300~fb^{-1}$ can put very stringent bounds: $|x_\gamma|< 1\times 10^{-3}$,
$|y_\gamma|< 4\times 10^{-3}$ and a strong correlation between the two.
\begin{figure}
\epsfxsize=8truecm
\centerline{\epsffile[50 263 506 693]{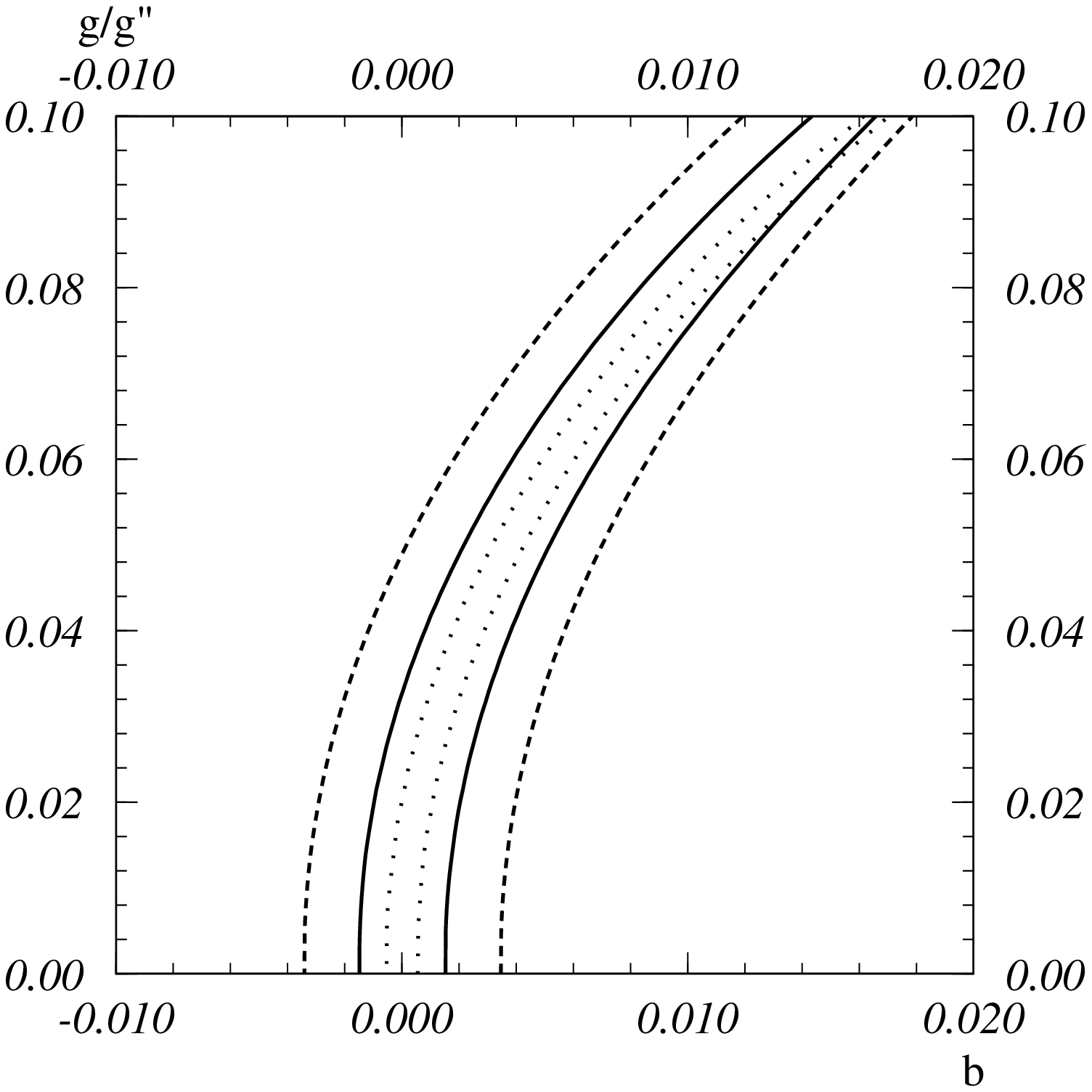}}
\caption{Same of Figure 3 for the parameter space $(b, g/g^{\prime\prime})$
in the case of  model B (BESS model).}
\end{figure}
\begin{figure}
\epsfxsize=8truecm
\centerline{\epsffile[50 263 506 693]{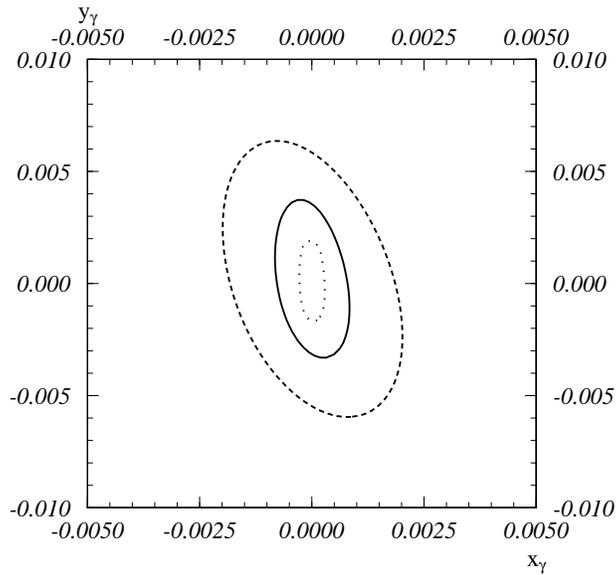}}
\caption{Same of Figure 3 for the parameter space $(x_\gamma, y_\gamma)$
in the case of  model C.}
\end{figure}

\noindent\underline{POLARIZED BEAMS}\\
\noindent
An interesting question is whether the use of polarized beams significantly
improves the bounds that can be placed on the anomalous couplings. In ref.
\cite{Dafne} we included in the analysis a longitudinal polarization for the
electron beam only. The result was encouraging since the general effect of the
polarization is to select different combinations of the anomalous parameters
producing, as a consequence, a rotation of the allowed region. Here
  we consider the possibility of longitudinally polarized electron and positron
beams. Depending on the mechanism used to polarize the beams it should be
possible to achieve various degrees of polarization. A reasonable possibility
\cite{ruckl}  seems to be $(P_L(e^-),P_L(e^+)) =  (0.9,0.6)$. We have considered
also the options with electron and positron beams polarized with opposite sign.
\begin{figure}
\epsfxsize=8truecm
\centerline{\epsffile[50 263 506 693]{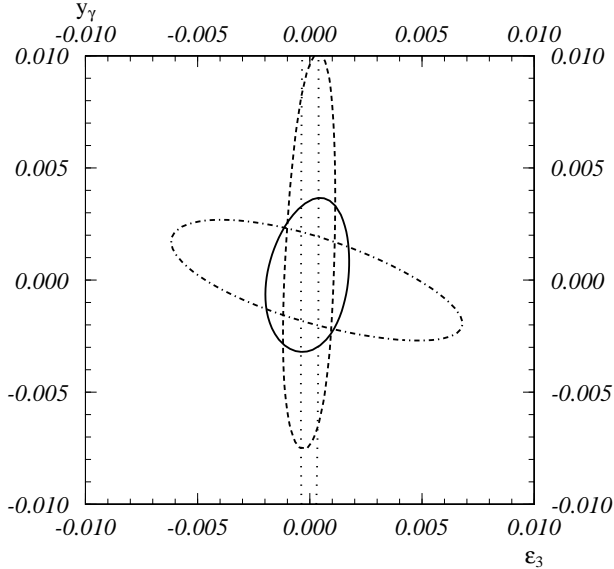}}
\caption{90\% CL bounds on the parameter space $(\epsilon_3,y_\gamma)$
in the case of model A. The bounds are obtained from $\sigma_{\rm comb}$ at
$E=500~GeV$, $L=300~fb^{-1}$ and
 different degrees of polarization $(\Pe,\Pep)$: the
 dash line corresponds to (0.9,0.6), the dot one  to (0.9,-0.6), the dash-dot to
 (-0.9,0.6) and the solid  one to (0,0).}
\end{figure}
\begin{figure}
\epsfxsize=8truecm
\centerline{\epsffile[50 263 506 693]{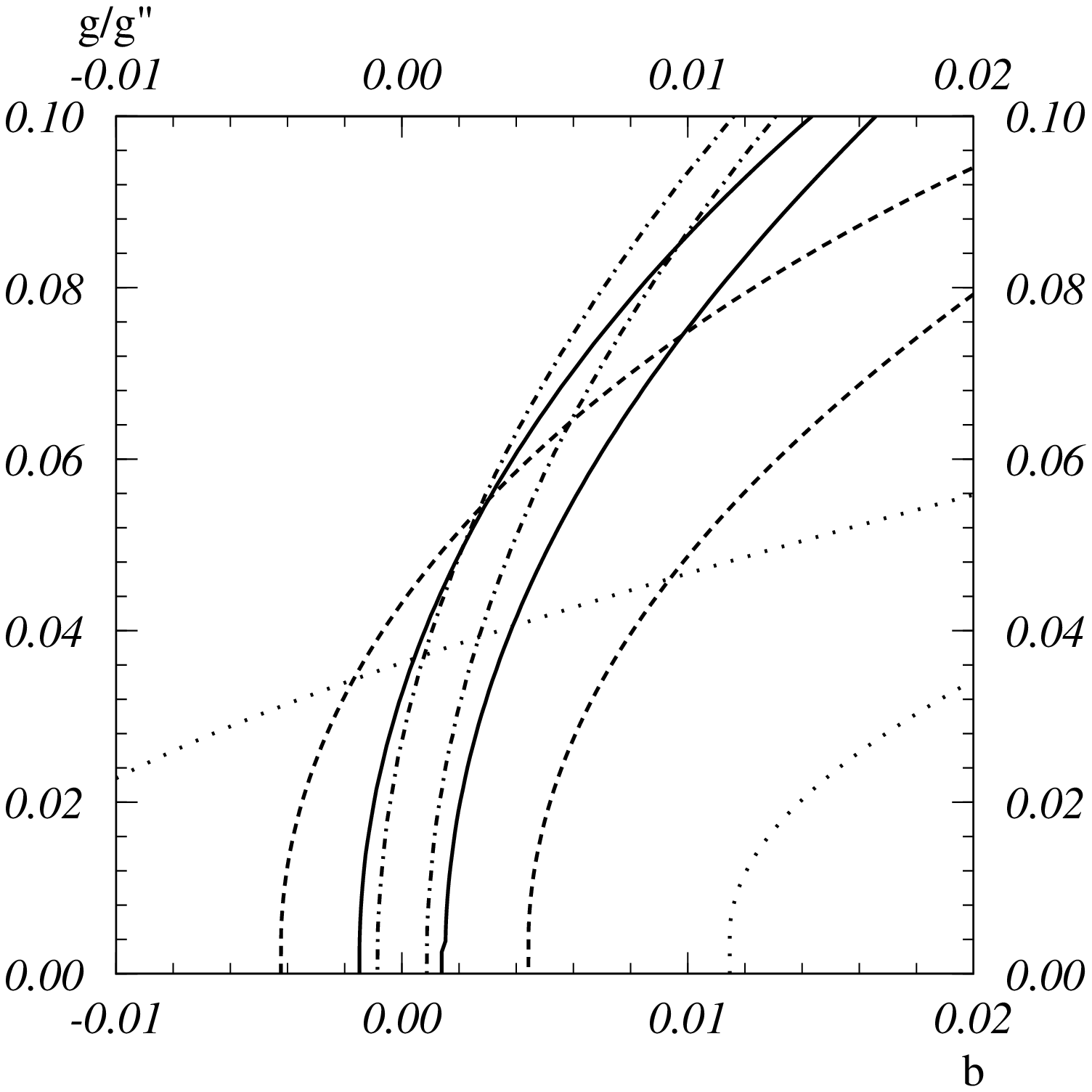}}
\caption{Same of Figure 6 for the parameter space $(b,g/g^{\prime\prime})$
in the case of  model B (BESS model).
  }\end{figure}
\begin{figure}
\epsfxsize=8truecm
\centerline{\epsffile[50 263 506 693]{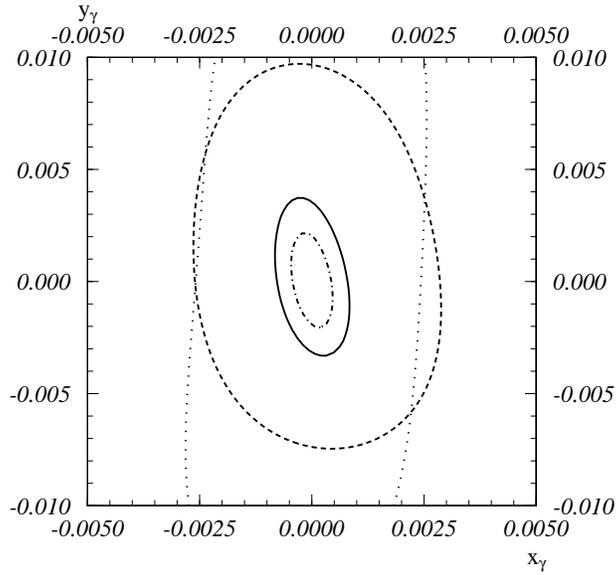}}
\caption{Same of Figure 6 for the parameter space $(x_\gamma, y_\gamma)$
in the case of  model C.}
\end{figure}
In Fig. 6 we give the 90\% CL bounds for the model A arising from combining the
final state polarized differential cross-sections at the energy of 500 $GeV$ and
luminosity of 300 $fb^{-1}$, for  different longitudinal degrees of polarization
of the incoming beams. We see that any degree of polarization leads to an
improvement of the bounds. However, the most efficient ones are the case with
$(P_L(e^-),P_L(e^+)) =  (\pm0.9,\mp0.6)$. For example, for this model an
increasing of the bounds on $y_\gamma$ with respect to the unpolarized case,
requires negative longitudinal polarization for the electrons while the best for
limiting $\epsilon_3$ is obtained with negative longitudinal polarization for
the positrons. We have not combined the bounds coming from different
polarizations, just to illustrate the achievements of the different  options. In
Fig. 7 the same analysis is performed for the BESS model. We see that, also in
this case,  it is much more efficient to consider polarizations with opposite
signs. A negatively polarized positron beam can put a very stringent bound on
the gauge coupling constant $g''$ of the new resonances. For the model C in Fig.
8 we plot the analogue of Figs.  6 and 7 and  we notice that, in this case, the
polarization of the initial beams does not give a bigger constraint except for
the case of a high degree of negative polarization for the incoming electrons.
\newline\indent
A general conclusion is   that polarization of the initial beams   is very
useful to constrain  the anomalous parameters. In addition, for any of the
options we have considered, we find that, by combining polarized and unpolarized
differential cross-sections, one is able to get significantly more restrictive
bounds, showing the complementarity of polarized and unpolarized measurements.

\resection{Conclusions}

In this contribution we have  considered the cross sections of $e^+e^-\to
W^+W^-$ in various channels of $W$'s polarizations as our observables to test
non-standard gauge and fermionic  couplings at the various options planned for
the Next Linear Colliders. We found that the longitudinal cross section gives
the most restrictive bounds on the anomalous parameters. We considered the
contribution of the anomalous fermionic couplings because they can be present in
a new physics scenario on the same footing of the trilinear anomalous gauge
couplings. Also, from the high-energy behaviour of the observables, it is easy
to show that terms proportional to the square of the anomalous couplings are not
negligible. So we performed all the calculation without making any expansion in
these parameters.
\newline\indent
We studied the effect of the polarization of the electron and positron beams to
the differential  cross sections. The general effect of the polarization is to
extend the reach of the colliders allowing experiments to probe different
directions in parameter space. In this way, combining bounds from different
polarization options, we obtain a much more restricted region for the anomalous
parameters. Our conclusion is that, for the particular models we have
considered, the most efficient option would be to have both beam polarized but
with opposite sign.
\newline\indent
 In summary we have seen that the NLC (already with a c.m. energy of 500 $GeV$
with the high-luminosity option) can give very stringent bounds on the anomalous
parameters also improving our knowledge of the fermionic couplings which are
very much constrained yet from LEP. Furthermore, if polarized beams will be
available, the bounds will increase dramatically. So the final result is that
polarized beams with adjustable degrees of polarization would constitute a very
significant tool in the search for new physics.
\newline\indent
Just to qualitatively compare these results for NLC  with the sensitivity to the
anomalous couplings of the present and planned accelerators we can refer to Fig.
5 of ref. \cite {barklow}. Pre-LHC experiments will not probe these couplings to
precision better than ${\cal O}(10^{-1})$. The present direct measurements of
$WWV$ couplings come from hadron collider experiments at Tevatron and from LEP2.
For example, the most restrictive 95\% CL limits on anomalous $WW\gamma$ and
$WWZ$ couplings from the $D0$  collaboration at Tevatron given in
\cite{tevatron}, assuming HISZ \cite{HISZ} equations (see eq. (\ref{g1z})), are
$-0.18\le \lambda_{\gamma}\le 0.19$ $(\Delta k_\gamma=0)$ and $-0.29\le\Delta
k_\gamma\le 0.53$         $(\lambda_\gamma=0)$ for $\Lambda=2~TeV$ (the momentum
scale at which in these couplings are calculate).
 Similar bounds come from LEP2 analysis. The results presented at the Tampere
conference \cite{tampere} gives 95\% CL one-parameter limits on $\Delta
k_\gamma$, $\lambda_\gamma$, $\Delta g_1^Z$ in the range 0.1-0.5. Limits will
improve at the Tevatron Upgrade and at the LHC. The analysis done at the LHC
shows that there is a strong dependence upon the assumed form factor scale
$\Lambda$. Anyway the various estimates are in the range $5-10 \times 10^{-3}$.
For the most recent analysis combining both CMS and ATLAS results, see
\cite{LHC}.
\newline\indent
As we have seen, a NLC collider at 500 $GeV$ c.m. energy would reach  ${\cal
O}(10^{-3})$ with unpolarized beams (if polarization will be available the
bounds could be pushed to ${\cal O}(10^{-4})$. This because the $e^+e^-$
colliders allow precision measurement of helicity amplitudes at well determined
c.m. energy and this is crucial to improve our understanding of the gauge boson
self-interactions and of the fermionic couplings.

\end{document}